\documentclass[iop,numberedappendix]{emulateapj-rtx4}
\usepackage{graphicx,natbib,color,bm,url,times}
\graphicspath{{./fig/}{./png/}}

\newcommand{\EQ}{\begin{equation}}
\newcommand{\EN}{\end{equation}}
\newcommand{\EQA}{\begin{eqnarray}}
\newcommand{\ENA}{\end{eqnarray}}

\newcommand{\EEq}[1]{Equation~(\ref{#1})}
\newcommand{\Eq}[1]{Equation~(\ref{#1})}
\newcommand{\Eqs}[2]{Equations~(\ref{#1}) and~(\ref{#2})}

\newcommand{\App}[1]{Appendix~\ref{#1}}
\newcommand{\Sec}[1]{Section~\ref{#1}}

\newcommand{\Fig}[1]{Figure~\ref{#1}}

\newcommand{\Figp}[2]{Figure~\ref{#1}({#2})}

\newcommand{\Figs}[2]{Figures~\ref{#1} and \ref{#2}}

\newcommand{\bra}[1]{\langle #1\rangle}

{}
{}
{}

{}
{}
{}
{}
{}
{}
{}
{}
{}
{}
{}
{}
{}
{}
{}
{}
{}

{}

\newcommand{\meanA}{\overline{A}}
\newcommand{\meanB}{\overline{B}}

\newcommand{\hatkk}{\hat{\bm{k}}}
\newcommand{\hatAA}{\hat{\bm{A}}}
\newcommand{\hatBB}{\hat{\bm{B}}}
{}

{}
{}
{}
\newcommand{\hatA}{\hat{A}}
\newcommand{\hatB}{\hat{B}}

\newcommand{\meanAA}{{\overline{\bm{A}}}}
\newcommand{\meanBB}{{\overline{\bm{B}}}}

\newcommand{\kk}{\bm{k}}
\newcommand{\KK}{\bm{K}}

\newcommand{\xx}{\bm{x}}
\newcommand{\XX}{\bm{X}}

\newcommand{\BB}{\bm{B}}

\newcommand{\JJ}{\bm{J}}
\newcommand{\oo}{\bm{\omega}}
\newcommand{\AAA}{\bm{A}}

\newcommand{\uu}{\bm{u}}

\newcommand{\nab}{{\bm{\nabla}}}

\newcommand{\ii}{{\rm i}}

\newcommand{\sgn}{{\rm sgn}  \, {}}

\newcommand{\dd}{{\rm d} {}}

\def\degr{\hbox{$^\circ$}}

\def\Pm{\mbox{\rm Pr}_M}
\def\Rm{R_{\rm m}}

\def\Rey{\mbox{\rm Re}}
\def\Imag{\mbox{\rm Im}}

\def\kf{k_{\rm f}}

\def\urms{u_{\rm rms}}

\def\half{{\textstyle{1\over2}}}

\def\quarter{{\textstyle{1\over4}}}

\newcommand{\G}{\,{\rm G}}

\newcommand{\m}{\,{\rm m}}

\newcommand{\Mm}{\,{\rm Mm}}
\newcommand{\Mx}{\,{\rm Mx}}

\newcommand{\J}{\,{\rm J}}
\newcommand{\yapj}[3]{ #1, {ApJ,} {#2}, #3}

\newcommand{\yapjl}[3]{ #1, {ApJ,} {#2}, #3}

\newcommand{\yan}[3]{ #1, {Astron.\ Nachr.,} {#2}, #3}

\newcommand{\yana}[3]{ #1, {A\&A,} {#2}, #3}

\newcommand{\ypf}[3]{ #1, {Phys.\ Fluids,} {#2}, #3}

\newcommand{\ypp}[3]{ #1, {Phys.\ Plasmas,} {#2}, #3}

\newcommand{\yprl}[3]{ #1, {Phys.\ Rev.\ Lett.,} {#2}, #3}

\newcommand{\ymn}[3]{ #1, {MNRAS,} {#2}, #3}
\newcommand{\ynat}[3]{ #1, {Nature,} {#2}, #3}

\newcommand{\ysph}[3]{ #1, {Solar Phys.,} {#2}, #3}

\newcommand{\yprd}[3]{ #1, {Phys.\ Rev.\ D,} {#2}, #3}
\newcommand{\ypre}[3]{ #1, {Phys.\ Rev.\ E,} {#2}, #3}

\newcommand{\yssr}[3]{ #1, {Spa.\ Sci.\ Rev.,} {#2}, #3}

\newcommand{\yjour}[4]{ #1, {#2}, {#3}, #4}

\newcommand{\ybook}[3]{ #1, {#2} (#3)}

\newcommand{\arxiv}[2]{ #1, arXiv:#2}
\def\apjs{ApJS}

\begin{document}
\title{Two-scale analysis of solar magnetic helicity}
\author{Axel Brandenburg$^{1,2,3,4}$, Gordon J. D. Petrie$^5$, \& Nishant K. Singh$^3$
}
\affil{
$^1$Laboratory for Atmospheric and Space Physics, University of Colorado, Boulder, CO 80303, USA\\
$^2$JILA and Department of Astrophysical and Planetary Sciences, University of Colorado, Boulder, CO 80303, USA\\
$^3$Nordita, KTH Royal Institute of Technology and Stockholm University, Roslagstullsbacken 23, SE-10691 Stockholm, Sweden\\
$^4$Department of Astronomy, AlbaNova University Center, Stockholm University, SE-10691 Stockholm, Sweden\\
$^5$National Solar Observatory, 3665 Discovery Drive, Boulder, CO 80303, USA
}
%\submitted{\today,~ $ 
%\submitted{$ $Revision: 1.128 $ $}
\submitted{Astrophys. J. 836, 21 (2017)}
\date{Received 2016 October 17; revised 2016 December 26; accepted 2016 December 31; published 2017 February 6}

\begin{abstract}
We develop a two-scale formalism to determine global magnetic helicity
spectra in systems where the local magnetic helicity has opposite signs
on both sides of the equator, giving rise to cancellation
with conventional methods.
We verify this approach using first a synthetic one-dimensional magnetic
field and then two-dimensional slices from a three-dimensional $\alpha$
effect-type dynamo-generated magnetic field with forced turbulence
of opposite helicity above and below the midplane of the domain.
We then apply this formalism to global solar synoptic vector magnetograms.
To improve the statistics, data from three consecutive Carrington rotations
(2161--2163) are combined into a single map.
We find that the spectral magnetic helicity representative of the northern
hemisphere is negative at all wavenumbers and peaks at $\approx0.06\Mm^{-1}$
(scales around $100\Mm$).
There is no evidence of bihelical magnetic fields that are found
in three-dimensional turbulence simulations of helicity-driven
$\alpha$ effect-type dynamos.
\end{abstract}

\email{brandenb@nordita.org}

\section{Introduction}

Magnetic helicity is a conserved quantity not only in ideal
magnetohydrodynamics (MHD), but also in nonideal MHD in the limit
of large magnetic Reynolds numbers.
It plays a crucial role in the theory of astrophysical
large-scale dynamos, given that in many cosmic bodies, the magnetic
Reynolds numbers are very large indeed.
Dynamo theory is relevant to explaining the global cyclic magnetic
field of the Sun \citep{KR82,GD96,B01,FB02}.
Magnetic helicity is also a topological invariant characterizing the
linkage of magnetic field lines and thus its complexity.
Large values of magnetic helicity appear to be connected with the
launching of coronal mass ejections \citep{Low94,NZZ03,Amari},
which in turn are relevant to understanding space weather.
This led to the quantitative evaluation of magnetic helicity in volumes
above the solar surface around active regions (ARs);
see \cite{PVDD15} for recent work.

To make contact with possible mechanisms that generate helical magnetic
fields, it is necessary to decompose magnetic helicity into contributions
from different length scales.
It has been known for some time that the $\alpha$ effect produces
a so-called bihelical magnetic field, with opposite signs at small and
large wavenumbers \citep{See96,Ji99,BB03}.
At each position, the net helicity integrated over contributions from
all scales is then actually zero.
Thus, to make meaningful quantitative statements, one needs to compute
magnetic helicity {\em spectra}; see \cite{YB03} for
examples of such fields produced by the $\alpha$ effect.

As a preliminary means of obtaining information about the magnetic
helicity of the large-scale field, one can use the azimuthally averaged
magnetic field to compute $2\meanA_\phi\meanB_\phi$, where $\meanA_\phi$
is the mean toroidal vector potential and $\meanB_\phi$ is
the mean toroidal magnetic field.
By taking the sign of $\meanB_\phi$ using Hale's polarity law and
computing $\meanA_\phi$ from the spherical harmonics decomposition of the
mean radial magnetic field, \cite{BBS03} concluded that, in the northern
hemisphere, $\meanA_\phi\meanB_\phi$ was negative (positive) before (after)
solar maximum.
A similar dependence was also obtained by \cite{ZSPGXSK10} by measuring
the current helicity.
Using synoptic vector magnetograms, \cite{PP14} computed $\meanA_\phi$
and $\meanB_\phi$ to obtain the global magnetic helicity of the
large-scale field of the Sun.
They found positive magnetic helicity in the north and negative in the
south, as expected from dynamo theory.

Magnetic helicity spectra are similar to magnetic energy spectra,
which have been computed for the Sun for some time
\citep{NP73,Abramenko05,Stenflo12}.
Magnetic helicity spectra can be computed analogously,
but this has only recently been attempted \citep{ZBS14,ZBS16}.
Those spectra can be of different signs in different
wavenumber ranges and at different positions on the solar surface.
Particularly important for the solar dynamo is the possibility
of a systematic dependence on solar latitude.
The question therefore arises as to how to analyze and present such
complex dependencies on position and scale in an efficient way.

A simple approach would be to determine spectra in different
local patches, but this can only be meaningful if the patches
are not too large.
This is indeed what has been done in the work of \cite{ZBS14,ZBS16},
who used patches of $(186\Mm)^2$.
However, a more elaborate technique has been developed in mean-field
dynamo theory by \cite{RS75} to separate large and small scales.
This is generally referred to as {\em two-scale analysis}.
It involves a so-called double Fourier transform, and allows
one to compute quadratic small-scale correlations such as the mean
electromotive force at large scales as a function of position based on
Fourier transforms of the constituent fields.
This is particularly important if the large length scales of interest
are not just the full spatial extent (corresponding to zero wavenumber),
but a somewhat smaller scale (finite wavenumber) on which physical
properties of the system vary slowly.
This is relevant to the Sun, where one can expect statistically similar
conditions at all longitudes, but only within broad bands in latitude.
Regarding the helicity of the magnetic field, for example, one expects
opposite signs in the northern and southern hemispheres
\citep[e.g.][]{PCSH08,PBNNvDG14}, so one would obtain zero
when averaging over north and south.

In this paper we begin by demonstrating the properties of the
double Fourier transform.
We consider first simple one-dimensional (1D) helical
and bihelical magnetic fields in the presence of an equator,
where the helicity changes sign.
Next, we apply the two-scale analysis to a three-dimensional (3D) turbulent
dynamo with periodic boundary conditions and an equator in the middle
that cuts the domain into two halves with opposite helicity of the
forcing function.
Finally, we employ full disk vector magnetograms from the
Helioseismic and Magnetic Imager (HMI) on board the {\em Solar Dynamics
Observatory} ({\em SDO}) to obtain all three magnetic field components
at the two-dimensional (2D) surface of the Sun, to compute magnetic
helicity spectra from a sequence of synoptic maps.

\section{Formalism}

\subsection{The usual magnetic energy and helicity spectra}
\label{UsualHelicitySpectra}

In a periodic Cartesian domain, the usual magnetic energy spectrum
is given by the integral over shells in wavenumber space of the
Fourier-transformed magnetic energy density,
\EQ
E_{\rm M}(k)=\int_{\Omega_D}\half\hatBB(\kk)\cdot\hatBB^\ast(\kk)\,
k^{D-1}\,\dd\Omega,
\EN
where $\dd\Omega$ is the surface differential in Fourier space
in $D$ dimensions.
In $D=3$ dimensions, the surface of a $D$-dimensional unit sphere
is $\Omega_3=4\pi$. In $D=2$ dimensions,
$\Omega_2=2\pi$ is the circumference of a unit circle, while in $D=1$
dimensions, $\Omega_1=2$ corresponds to the two end points of a line.
Here and in the following discussion, we measure the magnetic energy density
in $\G^2$ rather than $\J\m^{-3}$, so the vacuum permeability factor
is dropped.
Furthermore, $k=|\kk|$ is the radius of a sphere in Fourier space, and
hats denote the Fourier transform of the magnetic field, that is,
\EQ
\hatBB(\kk)=\int\BB(\xx)\,e^{-\ii\kk\cdot\xx}\,\dd^D\!x/(2\pi)^D.
\EN
The magnetic helicity spectrum is defined analogously to
$E_{\rm M}(k)$ as
\EQ
H_{\rm M}(k)=\int_{\Omega_D}\half\left[\hatAA\cdot\hatBB^\ast
+\hatAA^\ast\cdot\hatBB\right]\, k^{D-1}\,\dd\Omega,
\EN
where $\hatAA(\kk)$ is the Fourier transform of the vector potential
$\AAA(\xx)$ with $\nab\times\AAA=\BB$ and $\hatBB=\ii\kk\times\hatAA$.
These spectra are normalized such that
\EQA
&&\int_0^\infty E_{\rm M}(k)\,\dd k=\half\bra{\BB^2}_V\equiv{\cal E}_{\rm M},\\
&&\int_0^\infty H_{\rm M}(k)\,\dd k=\bra{\AAA\cdot\BB}_V\equiv{\cal H}_{\rm M},
\label{defEandH}
\ENA
%where angle brackets denote volume averages.
%AB: added with subscript $V$
where angle brackets with subscript $V$ denote volume averages.
Analogously, one can define the current helicity spectrum
$H_{\rm C}(k)$ such that $\int H_{\rm C}(k)\,\dd k=\bra{\JJ\cdot\BB}_V$,
where $\JJ=\nab\times\BB$ is proportional to the current density.

For the following, it is useful to remember that the magnetic energy spectrum
is the Fourier transform of the trace of the two-point correlation tensor
\EQ
M_{ij}(\xx)=\int\bra{B_i(\XX)B_j(\XX+\xx)}\,\dd^D\!X,
\label{Mxx}
\EN
where angle brackets denote ensemble averaging, which could be
approximated by averaging over time, and $M_{ij}(\xx)$ is assumed to be
statistically independent of $\XX$ owing to the assumption of homogeneity.
Thus, following standard relationships \citep[e.g.][]{MGS82,BSBG11},
the energy spectrum is then given by $2E_{\rm M}(k)=
\int \delta_{ij}\hat{M}_{ij}(\kk)\,k^{D-1}\,\dd\Omega$,
where $\hat{M}_{ij}(\kk)=\int M_{ij}(\xx)\,e^{-\ii\kk\cdot\xx}\,
\dd^D x/(2\pi)^D$ is the Fourier transform of $M_{ij}$.
Likewise, the scaled magnetic helicity spectrum is given by
$kH_{\rm M}(k)=\int\ii\hat{k}_i\epsilon_{ijk}\hat{M}_{jk}(\kk)\,k^{D-1}\,
\dd\Omega$, where $\hat{k}_i=k_i/|\kk|$ is the unit vector of $\kk$.
(The hat on $\kk$ is not to be confused with the hats on $\BB$ or
$M_{ij}$, where they denote Fourier transforms over $\kk$.)
(Including here the factors $2$ and $k$ on the left sides of the
two equations has the advantage that $2E_{\rm M}(k)$ and
$kH_{\rm M}(k)$ thus have the same prefactors.)
Note that under isotropic conditions, $k^2H_{\rm M}(k)=H_{\rm C}(k)$.
Such spectra show better the high-wavenumber range better than just
$H_{\rm M}(k)$ or $kH_{\rm M}(k)$ and are therefore also
considered in some of the following cases.

The magnetic helicity spectrum can be defined analogously from
a tensorial generalization of the magnetic two-point correlation tensor.
This will be done in the following section, where we will also
relax the assumption of homogeneity and will allow the spectra
to be slowly varying functions of $\XX$.

\subsection{Double Fourier transform}

To obtain magnetic energy and helicity spectra
that vary slowly in space, we consider
the two-point correlation tensor at position $\XX$ \citep{RS75},
\EQ
M_{ij}(\XX,\xx)=\bra{B_i(\XX+\half\xx)\,B_j(\XX-\half\xx)},
\EN
where $\xx$ is the distance between two points around $\XX$.
This expression is similar to \Eq{Mxx}, except that the
dependence on the slowly varying coordinate $\XX$ has been
retained and the two points of the two-point correlation
function are now symmetric about $\XX$.
In the following, we refer to such an analysis involving dependencies
on both $\xx$ and $\XX$ as a two-scale approach.
Sometimes we refer to this also as a global approach, as opposed to the
local approach involving smaller patches within just one hemisphere,
as done in \cite{ZBS14,ZBS16}.

By performing a Fourier transformation over $\xx$, we obtain
\EQ
\hat{M}_{ij}(\XX,\kk)=\int M_{ij}(\XX,\xx) \, e^{-\ii\kk\cdot\xx} \,
\dd^Dx/(2\pi)^D
\EN
for the spectral correlation tensor.
The symmetric part of this tensor contains information about the
energy spectrum, and the anti-symmetric components contain information
about the magnetic helicity spectrum.
Under isotropic conditions, $\hat{M}_{ij}(\XX,\kk)$ can be represented as
\EQ
\hat{M}_{ij}=[(\delta_{ij}-\hat{k}_i\hat{k}_j)\,2E_{\rm M}
-\ii\hat{k}_k\epsilon_{ijk}\,kH_{\rm M}]/2k^{D-1}\Omega_D,
\EN
and the magnetic energy spectrum is given by
\EQ
2E_{\rm M}(\XX,k)=\int_{\Omega_D}\delta_{ij}\hat{M}_{ij}(\XX,\kk)
\,k^{D-1}\,\dd\Omega,
\label{2EXk}
\EN
while the magnetic helicity spectrum (scaled with $k$) is given by
\citep{RS75}
\EQ
kH_{\rm M}(\XX,k)=\int_{\Omega_D}\ii\hat{k}_i\epsilon_{ijk} \hat{M}_{jk}(\XX,\kk)
\label{kHXk}
\,k^{D-1}\,\dd\Omega.
\EN
Except for the $\XX$ dependence, this formula is equivalent to that
used to estimate magnetic helicity in the solar wind \citep{MGS82,BSBG11}
and at the solar surface \citep{ZBS14,ZBS16}; see also the end of
\Sec{UsualHelicitySpectra}.

In practice, we are interested in the case where the helicity varies
in latitude and changes sign at the equator.
Before discussing the magnetic field at the surface of the Sun, where
the field is given in spherical coordinates, we consider examples in
Cartesian coordinates.
This is technically and conceptionally easier.
In that case we consider a cubic domain of size $L^3$, which is homogeneous
in the $x$ and $y$ directions, but inhomogeneous in the $z$ direction, 
such that $0\leq z\leq\pi$ corresponds to the northern hemisphere
and $-\pi\leq z\leq0$ to the southern.
Thus, we are interested in slow changes of the magnetic energy and
helicity spectra as a function of $\XX=(X,Y,Z)$, where we shall be
specifically interested in the dependence on $Z$, which corresponds
to latitude or distance from the equator or the midplane.
This can easily be done by performing an additional Fourier transform over
$\XX$:
\EQ
\tilde{M}_{ij}(\KK,\kk)=\int \hat{M}_{ij}(\XX,\kk) \, e^{-\ii\KK\cdot\XX} \,
\dd^DX/(2\pi)^D.
\EN
It can then be shown that \citep{RS75}
\EQ
\tilde{M}_{ij}(\KK,\kk)=\bra{\hatB_i(\kk+\half\KK) \,
\hatB_j^\ast(\kk-\half\KK)}.
\label{RS75kK}
\EN
Thus, slow variations of the spectrum correspond to a shift between two
points in wavenumber space.
By integrating again over ``shells'' in $\kk$ space, we obtain
$\KK$-dependent magnetic energy and helicity spectra analogously
to \Eqs{2EXk}{kHXk} as
\EQ
2\tilde{E}_{\rm M}(\KK,k)=\int_{\Omega_D}\delta_{ij}\tilde{M}_{ij}(\KK,\kk)
\,k^{D-1}\,\dd\Omega,
\label{2EKk}
\EN
\EQ
k\tilde{H}_{\rm M}(\KK,k)=\int_{\Omega_D}\ii\hat{k}_i\epsilon_{ijk} \tilde{M}_{jk}(\KK,\kk)
\label{kHKk}
\,k^{D-1}\,\dd\Omega.
\EN
Thus, the spectrum of magnetic helicity with a slow variation in
the $z$ direction is proportional to $\sin K_Z Z$ and is given by
$\KK=(0,0,K_Z)$, where $K_Z=2\pi/L$ and $z=Z$ are used interchangeably.

Unlike $H_{\rm M}(\XX,k)$, which is real, $\tilde{H}_{\rm M}(\KK,k)$
is complex.
The quantity of interest depends on the spatial profile of the
background helicity.
For the rest of this paper, we are concerned with helicity profiles
proportional $\sin K_0 Z$ with an equator at $Z=0$.
Its Fourier transform is $-\half\ii\delta(K_Z-K_0)$.
We will therefore plot the {\em negative imaginary part} of
$\tilde{H}_{\rm M}(\KK,k)$, which reflects the sign of
magnetic helicity in the northern hemisphere.

\section{Testing the formalism}

To verify that the two-scale formalism allows us to disentangle
the proper magnetic helicity from measurements over the full domain,
and thus both hemispheres, we apply it first to data where we know
the result: (i) a synthetically constructed 1D helical
Beltrami-like magnetic field and (ii) a 3D field from
a turbulent dynamo with a hemispheric modulation of the helicity of
the forcing function.

\subsection{A 1D example}
\label{A1Dexample}

A simple static 1D helical magnetic field is a Beltrami field
of the form $\BB=(\sin k_1z,\cos k_1z,0)$, but its magnetic helicity
density is uniform, because in this example, the vector potential is
parallel to $\BB$ with $\AAA=\BB/k_1$.
Here, $k_1$ is the wavenumber of the magnetic field, and the helicity is
positive for $k_1>0$ and is associated with a $\pi/2$ phase shift where
$B_y$ precedes $B_x$ as a function of $z$ by a phase shift of $\pi/2$.

\begin{figure}[t!]\begin{center}
\includegraphics[width=\columnwidth]{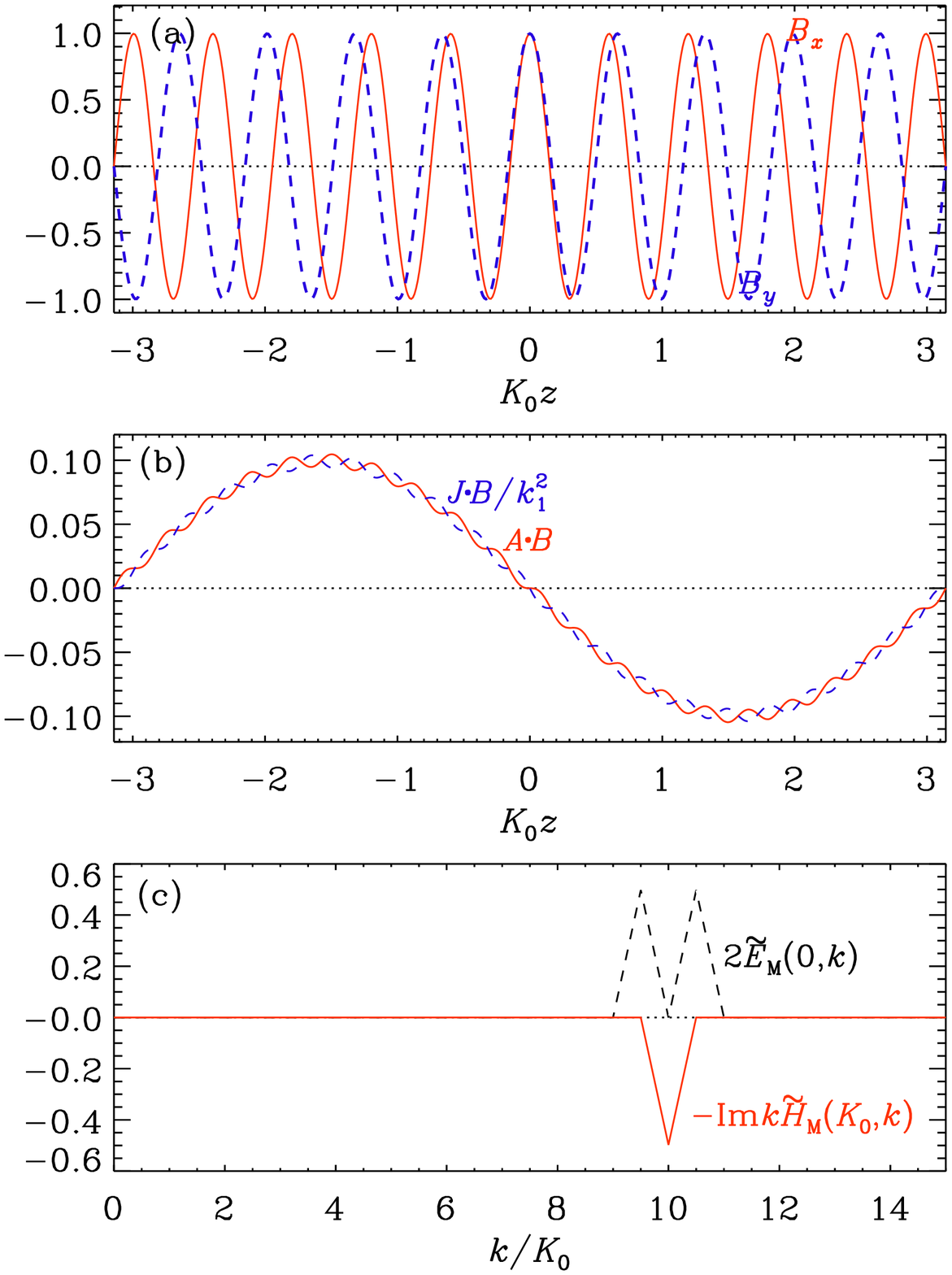}
\end{center}\caption[]{
(a) $B_x(z)$ and $B_y(z)$ from \Eq{singly2} with $k_1/K_0=-10$,
(b) $\AAA\cdot\BB$ and $\JJ\cdot\BB/k_1^2$, as well as
(c) $2\tilde{E}_{\rm M}(0,k)$ and $k\tilde{H}_{\rm M}(K_0,k)$.
}\label{psingle3}\end{figure}

To make the helicity density a slowly varying function of $z$,
the phase shift between $B_x$ and $B_y$ must also slowly change.
This is accomplished by having slightly different wavelengths for
$B_x$ and $B_y$.
\EEq{RS75kK} suggests the following form
\EQ
\BB=\left(\cos k_1^-z,\cos k_1^+z,0\right),
\label{singly2}
\EN
where $k_1^\pm=k_1\pm\half K_0$ and $K_0$ is the wavenumber of the
slowly varying magnetic helicity density; see \Figp{psingle3}{a}.
The corresponding vector potential is 
$\AAA=((k_1^+)^{-1}\sin k_1^+z,\;-(k_1^-)^{-1}\sin k_1^-z,\;0)$, so the
magnetic helicity density is then obtained as
\EQ
\AAA\cdot\BB=\left(k_1\sin K_0 z-\half K_0\sin2k_1 z\right)
/(k_1^2-\quarter K_0^2).
\EN
For $k_1\gg K_0$, we have $\AAA\cdot\BB\approx k_1^{-1}\sin K_0 z$;
see also \Figp{psingle3}{b}.
As shown in detail in \App{Deriv}, the magnetic energy spectrum
at $K=0$ is
\EQ
2\tilde{E}_{\rm M}(0,k)=\half\left(\delta_{k\,k_1^+}+\delta_{k\,k_1^-}\right)
\label{2EMK0kz}
\EN
and the magnetic helicity spectrum at $K=K_0$ is
\EQ
-\Imag k\tilde{H}_{\rm M}(K_0,k)=\half\sgn k_1 \; \delta_{k\,k_1},
\label{kHMK0kz}
\EN
see \Figp{psingle3}{c}.
Thus, the presence of small-scale helicity with a $K=K_0$
modulation is perfectly captured by the two-scale analysis.
In particular, the sign of $-\Imag k\tilde{H}_{\rm M}(K_0,k)$
is equal to the sign of the magnetic helicity in the northern hemisphere,
and thus equal to the sign of $k_1$.

\begin{figure}[t!]\begin{center}
\includegraphics[width=\columnwidth]{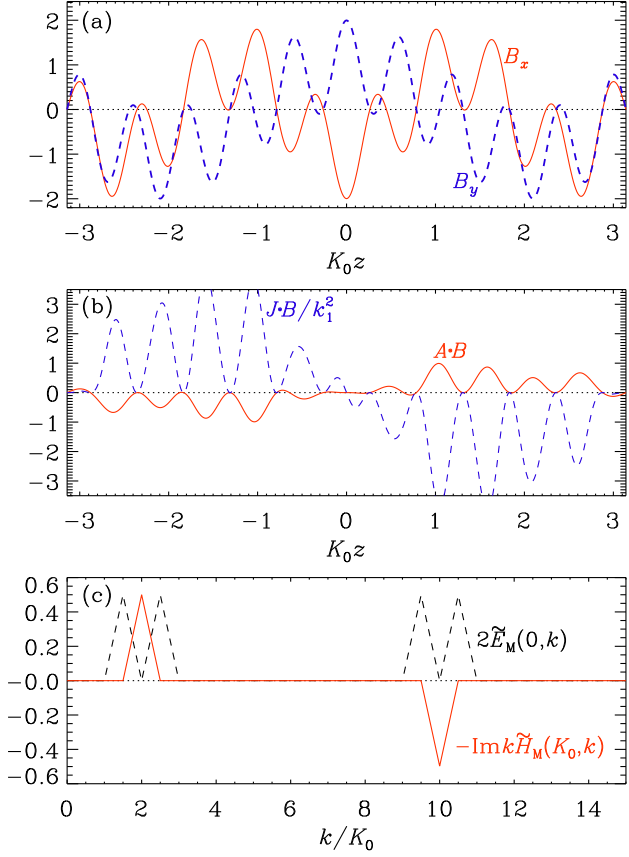}
\end{center}\caption[]{
Similar to \Fig{psingle3}, but for a bihelical field with
$k_1/K_0=2$ and $k_2/K_0=-10$.
}\label{pdouble3}\end{figure}

Next, we demonstrate in \Fig{pdouble3} that even a bihelical magnetic
field consisting of a superposition of two helical magnetic fields with
different wavenumbers and opposite signs can still easily be disentangled,
even though the helicities of both components are already modulated
and change sign proportional to $\sin K_0 z$.
In this case, the spatial profiles of $\AAA\cdot\BB$ and $\JJ\cdot\BB$
are no longer related to each other in a simple way and tend to have
opposite signs; see \Figp{pdouble3}{b}.

\subsection{3D turbulence}

\begin{figure}[t!]\begin{center}
\includegraphics[width=\columnwidth]{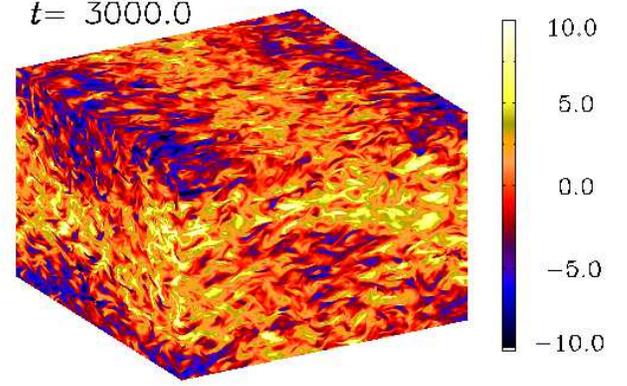}
\end{center}\caption[]{
Visualization of $B_x$ on the periphery of the domain.
The $z$ direction points upward, and the equatorial plane
is in the middle of the domain.
Note the occurrence of a large-scale magnetic field patches
together with a small-scale field on the scale of the forcing.
}\label{Bx3000}\end{figure}

We now consider the results for helically forced 3D
hydromagnetic turbulence in triply periodic domains.
The forcing is applied at a length scale that is ten times smaller
than the computational domain, so that large-scale dynamo action
on the scale of the domain is possible.
The helicity of the forcing function is proportional to $\sin K_0z$,
where $K_0=2\pi/L$ is the smallest wavenumber in our domain of size $L^3$.
To illustrate the appearance of such a field, we show in \Fig{Bx3000}
the component $B_x$ on the periphery of the domain for a simulation
with $576^3$ meshpoints at a magnetic Reynolds number,
$\Rm=\urms/\eta\kf\approx100$, where $\urms$ is the rms velocity of the
turbulence, $\eta$ is the magnetic diffusivity, and $\kf$ is the forcing
wavenumber with $\kf\approx10\,K_0$.
The magnetic Prandtl number is $\Pm=\nu/\eta=1$, where $\nu$ is the
kinetic viscosity.
The setup of these simulations is similar to that of \cite{MTKB10},
who found that such Cartesian dynamos with an equator produce helical
magnetic fields with equatorward migration, if the magnetic boundaries
in the $z$ direction are perfect conductors.
In the present case, on the other hand, a periodic boundary condition is used.
The present simulation has been performed with the
{\sc Pencil Code}\footnote{\url{https://github.com/pencil-code}}.

\begin{figure}[t!]\begin{center}
\includegraphics[width=\columnwidth]{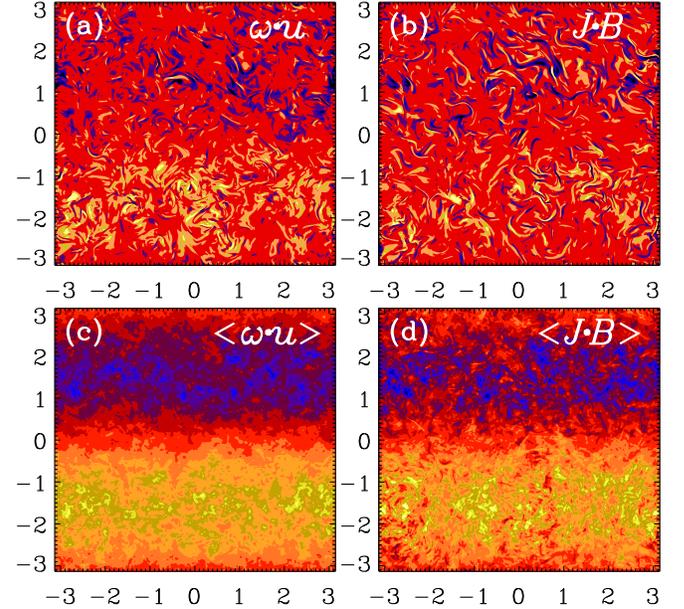}
\end{center}\caption[]{
Slices (upper row) and averages (lower row)
of $\oo\cdot\uu$ (left) and $\JJ\cdot\BB$ (right).
As in \Fig{Bx3000}, blue (yellow) denotes negative (positive) values.
}\label{ppjb}\end{figure}

\begin{figure}[t!]\begin{center}
\includegraphics[width=\columnwidth]{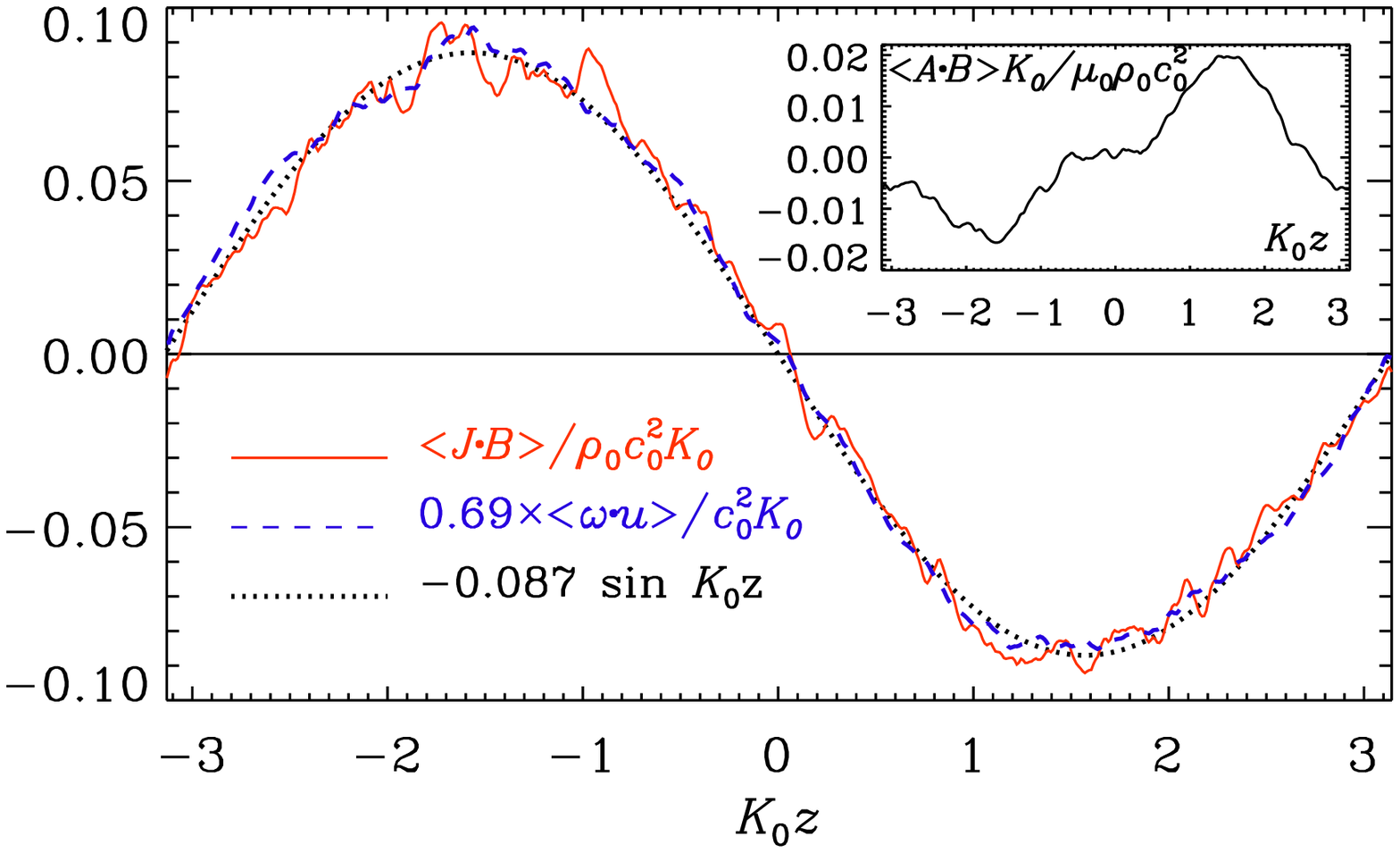}
\end{center}\caption[]{
Dependence of $\bra{\JJ\cdot\BB}$ and $\bra{\oo\cdot\uu}$ on $z$.
The inset shows $\bra{\AAA\cdot\BB}$.
}\label{ppjbm}\end{figure}

In \Fig{ppjb} we show $xz$ slices and averages of both kinetic and current
helicities (i.e., $\oo\cdot\uu$ with $\oo=\nab\times\uu$ being the
vorticity, and $\JJ\cdot\BB$, respectively).
The averages are taken along the direction normal to the plane.
The four panels show that both kinetic and current helicities are
negative above the midplane ($0<z<\pi$) and positive below ($-\pi<z<0$).
However, the individual slices show considerable fluctuations and
violations of the hemispheric sign rule, even though this simulation
has maximally helical forcing at $\sin K_0 z=1$ or $-1$.
The averages over the normal direction are much less noisy, although
fluctuations on the forcing scale can still be discerned.
In \Fig{ppjbm} we show $\bra{\oo\cdot\uu}_{xy}(z)$ and
$\bra{\JJ\cdot\BB}_{xy}(z)$,
i.e., where we have also averaged over the $x$ direction.
We clearly see the sinusoidal variation of the two mean helicity
densities, just like in panel (b) of \Figs{psingle3}{pdouble3}.
The inset of \Fig{ppjbm} shows that $\bra{\AAA\cdot\BB}_{xy}(z)$,
which is dominated by the large-scale field \citep{B01}, has
(as expected) the opposite sign.

\begin{figure}[t!]\begin{center}
\includegraphics[width=\columnwidth]{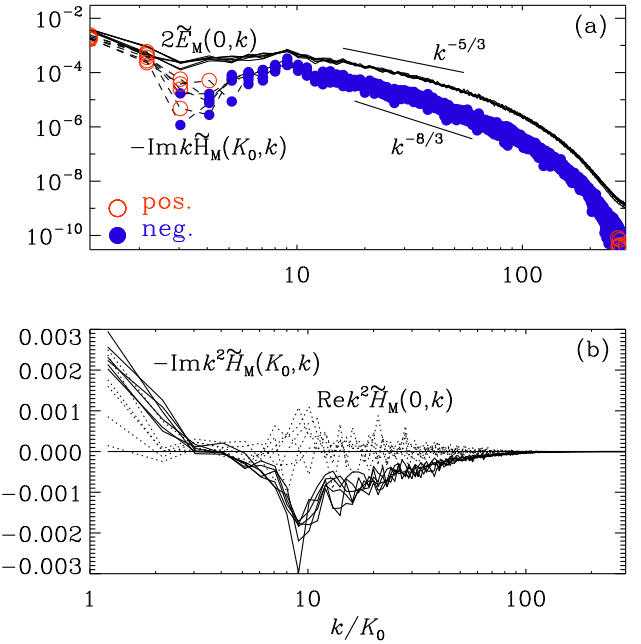}
\end{center}\caption[]{
(a) Magnetic energy and helicity spectra from six 2D slices
for 3D turbulence.
(b) Comparison between $-\Imag k^2\tilde{H}_{\rm M}(K_0,k)$
(solid lines) and $\Rey k^2H_{\rm M}(0,k)$ (dotted lines).
}\label{pslice_spectra}\end{figure}

To verify that the bihelical signature of dynamo-generated magnetic fields
can clearly be extracted from single slices of the entire volume from
simulated magnetic field data, we plot in \Fig{pslice_spectra}
magnetic energy and helicity spectra.
In panel (a) we plot the usual ($\KK=0$) magnetic energy spectrum,
$2\tilde{E}_{\rm M}(0,k)$, which shows the energy injection
scale at $k/K_0=10$ as well as magnetic energy on the scale
of the domain at $k/K_0=1$ and $2$.
The scaled magnetic helicity spectrum $-\Imag k\tilde{H}_{\rm M}(K,k)$
for $K=K_0$ can be positive and negative, so we plot
$|\Imag k\tilde{H}_{\rm M}(K_0,k)|$,
but indicate the two signs using different plot symbols.

At wavenumbers above the injection wavenumber, the magnetic energy
spectrum shows an approximate $k^{-5/3}$ subrange.
Furthermore, $|\Imag k\tilde{H}_{\rm M}|$ shows a $k^{-8/3}$ subrange,
which corresponds to a $k^{-5/3}$ spectrum for the current helicity,
as has been found previously using the usual helicity spectra in
fully homogeneous turbulence \citep{BS05b,B09}.
This variation should therefore be well suited for analysis with
the two-scale approach.
Note also that $|\Imag k\tilde{H}_{\rm M}|$ reaches peak values
at around 0.003, which is somewhat below the typical value of
$|\bra{\AAA\cdot\BB}_{xy}(z)|$; see the inset of \Fig{ppjbm}.

In \Figp{pslice_spectra}{b} we compare current helicity spectra
for $K=K_0$ and $K=0$, i.e., $-\Imag k^2\tilde{H}_{\rm M}(K_0,k)$
and $\Rey k^2\tilde{H}_{\rm M}(0,k)$, respectively,
computed for six uniformly separated horizontal planes.
Note that for the latter, the contributions from the planes
above and below $z=0$ tend to cancel and fluctuate around zero.
The $k^2$ factor has been applied to show more clearly the
relative strengths of the two extrema.

\section{Application to solar vector magnetograms}

We combine synoptic vector magnetogram from three successive
Carrington rotations (CRs).
Synoptic vector magnetograms, based on full-disk vector magnetograms from
the synoptic vector magnetograms from the Synoptic Optical Long-term
Investigations of the Sun (SOLIS) vector spectromagnetograph (VSM),
were first presented by \citet{Gosain13}.
Here we consider similar but higher-resolution ($0.1\degr$ at the
equator) synoptic maps for CR~2161--2163, constructed using full-disk vector
magnetograms obtained from {\em SDO}/HMI. The data were processed by Yang
Liu\footnote{\url{http://hmi.stanford.edu/hminuggets/?p=1689}} (Stanford).
It is worth noting that similar data, accompanied by uncertainty maps
but processed at lower intermediate resolution ($1\degr$),
are also available \citep{Hughes16}.

The magnetic field vector is expressed in spherical coordinates,
$(B_r,B_\theta,B_\phi)$, where $(r,\theta,\phi)$ correspond to
radius $r$, colatitude $\theta$, and longitude $\phi$.
We map the field onto the $(\phi,\mu)$ plane, where $\mu=\cos\theta$
increases from south to north.
This allows us to adopt a Cartesian analysis by substituting
\EQ
(\phi,\mu)\to(y,z),\quad (B_r,B_\phi,-B_\theta)\to(B_x,B_y,B_z).
\EN
This mapping preserves the right-handedness of the coordinate system.
We regard this approach as a substitute to what should ultimately be done
in spherical harmonics, but that would be technically rather different
from the previous test cases and will therefore be avoided here.
Furthermore, the use of $\mu$ instead of $\theta$ is not rigorously
justified, but it seems useful because it does de-emphasize in a natural
way data from high latitudes that are more uncertain.

\subsection{Local analysis}

\begin{figure}[t!]\begin{center}
\includegraphics[width=\columnwidth]{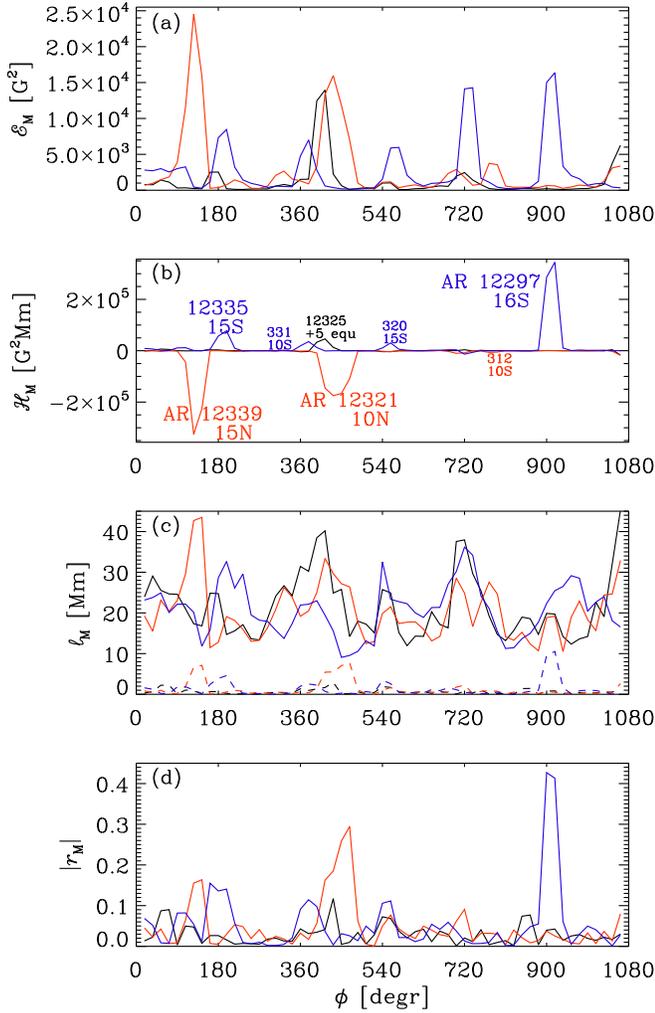}
\end{center}\caption[]{
(a) ${\cal E}_{\rm M}$, (b) ${\cal H}_{\rm M}$,
(c) $\ell_{\rm M}$ (solid lines) and
$|{\cal H}_{\rm M}|/2{\cal E}_{\rm M}$ (dashed lines),
and (d) $|r_{\rm M}|$ in three strips
between $5\degr$ and $35\degr$ northern latitude (N, red lines),
between $5\degr$ and $35\degr$ southern latitude (S, blue lines),
and at the equator between $\pm7\degr$ latitude (E, black lines) 
for CR~2161--2163 covering the period from 2015 February 28
(right) to 2015 May 20 (left).
}\label{pHHmNS_roll_allk}\end{figure}

Before applying the global two-scale analysis, we present in
\Fig{pHHmNS_roll_allk} the time evolution of ${\cal E}_{\rm M}$ and
${\cal H}_{\rm M}$, defined after \Eq{defEandH}, by computing the usual
magnetic helicity spectra for a sequence of 60 overlapping patches
with a width of $36\degr$ in longitude along three strips between
$5\degr$ and $35\degr$ north (N), $5\degr$ and $35\degr$ south (S),
and an equatorial strip between $\pm7\degr$ latitude.
Here, $0\degr\leq\phi\leq360\degr$ refers to CR~2163,
$360\degr\leq\phi\leq720\degr$ to CR~2162, and
$720\degr\leq\phi\leq1080\degr$ to CR~2161.

\begin{figure}[t!]\begin{center}
\includegraphics[width=\columnwidth]{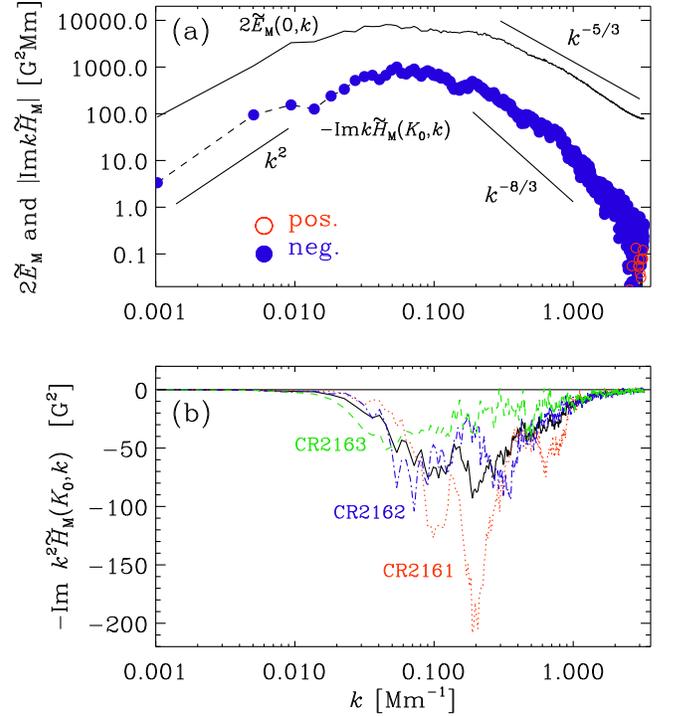}
\end{center}\caption[]{
(a) Magnetic energy and helicity spectra for the 2D solar surface
for the interval spanning CR~2161--2163.
(b) Current helicity spectrum, $-\Imag k^2\tilde{H}_{\rm M}(K_0,k)$,
for the interval spanning CR~2161--2163 (solid black) together with
the results for CR~2161 (dotted red), 2162 (dash-dotted blue),
and 2163 (dashed green).
}\label{pphelspec_two}\end{figure}

In \Figp{pHHmNS_roll_allk}{c} we also plot the evolution of the
so-called integral scale of the turbulence,
\EQ
\ell_{\rm M}=\left.\int_0^\infty k^{-1} \tilde{E}_{\rm M}(0,k)\,\dd k\,
\right/\int_0^\infty\tilde{E}_{\rm M}(0,k)\,\dd k
\EN
and compare with $|{\cal H}_{\rm M}|/2{\cal E}_{\rm M}\leq\ell_{\rm M}$,
which is known as the realizability condition \citep{KTBN13}.
The modulus of their ratio, $r_{\rm M}={\cal H}_{\rm M}/
2\ell_{\rm M}{\cal E}_{\rm M}$, can reach values of the order of 0.15--0.4;
see \Figp{pHHmNS_roll_allk}{d}.

It turns out that in most of the patches, the mean magnetic helicity
is very small, and large values are confined to just a few isolated
patches with strong ARs, where the magnetic helicity is large;
see \Figp{pHHmNS_roll_allk}{b}.
There are two strong ARs in the northern hemisphere (AR~12321 and AR~12339)
and one strong one in the southern hemisphere during CR~2161, namely AR~12297.
The magnetic helicity of all the other ARs is fairly weak, although there
are many occasions where there are prominent ARs.

\subsection{Global analysis}

Next, we consider magnetic helicity spectra obtained in the
two-scale approach.
In the Cartesian approach described above, the calculation of
$kH_{\rm M}(\KK,k)$ is straightforward.
We consider $kH_{\rm M}(\KK,k)$ versus $k$ for a fixed vector
$\KK=(0,K_0)$, i.e., we assume that there is no systematic modulation
in longitude and that $K_0=2\pi/2R_\odot=\pi/R_\odot$ is the projected
range for $-1\leq\mu\leq1$, where $R_\odot$ is the solar radius.
In the following, we therefore write for simplicity $H_{\rm M}(K,k)$,
i.e., with a scalar $K$.
For the energy spectrum, we consider, as before,
no modulation and thus just $2E(0,k)$ versus $k$.
The highest wavenumber corresponding to the resolution of $0.1\degr$
(or $1.2\Mm$) is $k_{\max}=1800/R_\odot\approx2.6\Mm^{-1}$.

As explained in \Sec{A1Dexample}, with the equator being at $\mu=0$, the
relevant quantity in this case is $-\Imag k\tilde{H}_{\rm M}(K_0,k)$.
It turns out that it is negative for almost all values of $k$;
see \Figp{pphelspec_two}{a}.
This is surprising and quite different from the corresponding
result for a helically driven large-scale dynamo.
Nevertheless, there are considerable variations in the value of the
spectrum if one compares with data from only one of any of the three CRs.
Those results are also shown in \Figp{pphelspec_two}{b}.
The range of variation can be regarded as an estimate of the error ``bar''
of $-\Imag k\tilde{H}_{\rm M}(K_0,k)$.
However, even though this quantity can change by a factor of $2$,
the sign still does not change.

\begin{figure}[t!]\begin{center}
\includegraphics[width=\columnwidth]{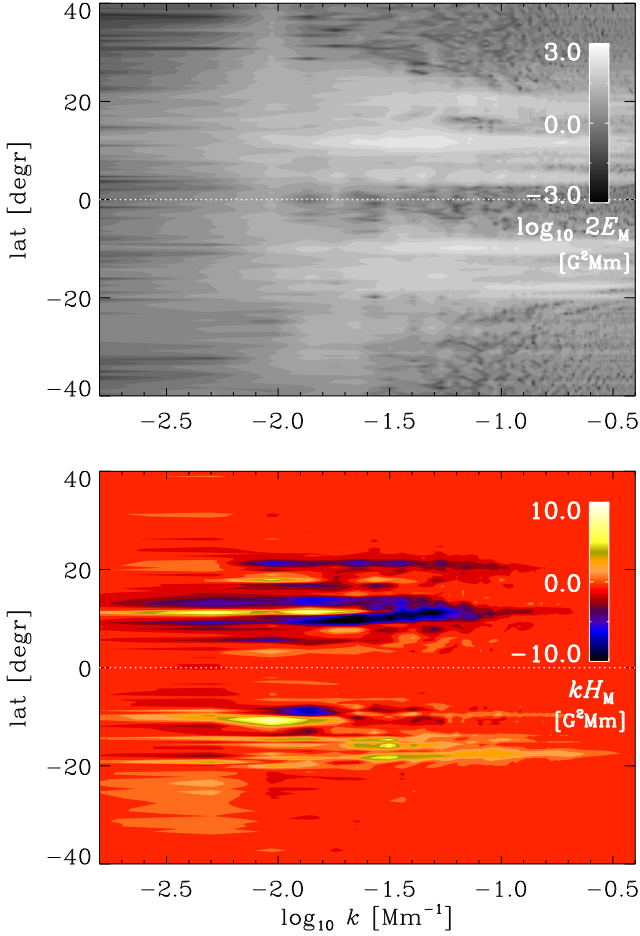}
\end{center}\caption[]{
Magnetic energy and helicity spectra for 2D solar surface data
for CR~2161--2163 as a function of $k$ and latitude, $\arcsin Z$.
}\label{pkHkNS_fullk}\end{figure}

If $K_0=0$, we can apply the realizability condition,
$|k\tilde{H}_{\rm M}(0,k)|\leq2\tilde{E}_{\rm M}(0,k)$.
By contrast, for $K_0\neq0$, this no longer holds,
so the separation between the graphs of $|\Imag k\tilde{H}_{\rm M}(K_0,k)|$
and $2\tilde{E}_{\rm M}(0,k)$ can no longer be used as quantitative measure
of the fractional magnetic helicity and how close it is to the maximum
possible value.

\subsection{Latitudinal dependence}

Finally, let us consider the latitudinal dependence, $\XX=(0,Z)$, at the
solar surface.
To do this, we have to transform back from $K$-space to $Z$-space and
then plot the spectra as a function of $Z$ (or $z$, or even $\mu$, which
are all equivalent).
The result is shown in \Fig{pkHkNS_fullk}, where we have computed the
return transformation as
\EQ
H_{\rm M}(Z,k)=\int e^{\ii K_Z Z}\tilde{H}_{\rm M}(K_Z,k)\,\dd K_Z.
\EN
We have computed the return transformation for $E_{\rm M}(Z,k)$ analogously.
Here, the Fourier integral has been evaluated as a Fast Fourier Transform
with $-128\leq K/K_0\leq127$, resulting in 256 points in $-1<\mu<1$.
It turns out that both $kH_{\rm M}(Z,k)$ and $2E_{\rm M}(Z,k)$ are
strongly concentrated along narrow latitudinal strips at $\pm15\degr$
latitude.
Again, the spectra are concentrated within the range
$0.01\Mm^{-1}\leq k\leq0.1\Mm^{-1}$.
As expected, the magnetic helicity is negative in the north, which is
consistent with the two-scale analysis, where a negative value of
$\Imag\tilde{H}(K_0,k)$ corresponds to negative magnetic helicity
in the north and positive in the south.

Let us emphasize at this point that by going into Fourier space,
we have automatically eliminated the gauge dependence of magnetic helicity.
This, in turn, is a consequence of the implicit assumption that the
input to the Fourier transform is periodic.
This assumption might reasonably well be justified if the domain extends
between both poles, where the field is weak anyway, and in longitude, if
sufficiently many synoptic maps are ``stitched'' together.
Ultimately, of course, the Fourier formalism should be replaced by
one involving spherical harmonics, similar to what has been done
previously for the mean magnetic field \citep{BBS03,PP14}.
However, this has not yet been developed in the context of the
two-scale formalism.

\subsection{Comparison with the azimuthally averaged mean field}

Let us now compare with the magnetic helicity density from the
azimuthally averaged mean magnetic field,
$\meanBB=\int_0^{2\pi}\BB\,\dd\phi/2\pi$, for which
the gauge-invariant relative magnetic helicity is given by
$2\int\meanA_\phi\meanB_\phi\,\dd^D\!x$ \citep{BDS02},
where $\meanA_\phi$ and $\meanB_\phi$ are the toroidal components
of $\meanAA$ and $\meanBB=\nab\times\meanAA$, respectively,
and $\meanA_\phi=\meanB_\phi=0$ on the axis.
We now compute the magnetic helicity density, $2\meanA_\phi\meanB_\phi$,
where $\meanA_\phi$ is related to
$\meanB_r=-\partial(\sin\theta\meanA_\phi)/\partial\mu$.
Analogously, in our Cartesian mapping, we have
$\meanB_x=-\partial\meanA_y/\partial z$.

\begin{figure}[t!]\begin{center}
\includegraphics[width=\columnwidth]{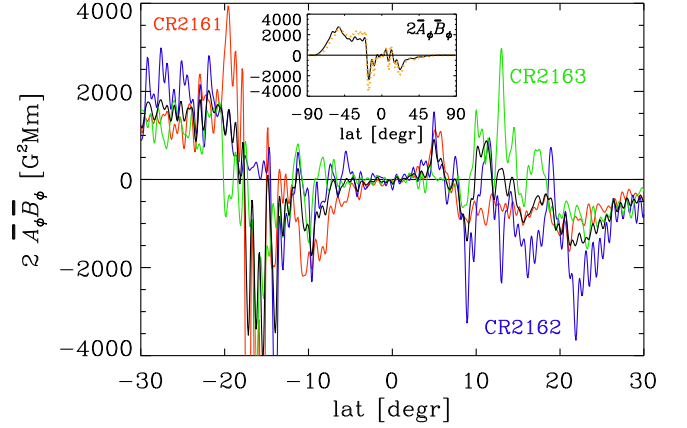}
\end{center}\caption[]{
Latitudinal dependence of $2\meanA_\phi\meanB_\phi$ for CR~2161--2163
(black), together with the results for the three CRs separately.
The inset shows the full range from pole to pole (black) and a comparison
with the result from the Cartesian analysis (orange dashed).
}\label{pcalc_bl}\end{figure}

To compute $\meanA_y$, it is convenient to employ spectral space,
i.e., $\hatA_y=-\hatB_x/\ii k_z$ in our Cartesian mapping.
Alternatively, in spherical coordinates, owing to axisymmetry,
we have \citep{BBS03,PP14}
\begin{equation}
\meanA_\phi(\mu)=-R\sum_{\ell=1}^{N_\ell} 
\frac{\ell+1/2}{\ell(\ell+1)}\hatB_\ell P_\ell^1(\mu),
\end{equation}
where $P_\ell^1(\mu)$ are the associated Legendre polynomials of
degree $\ell$ and order one,
$\hatB_\ell=\int_{-1}^1 \meanB_r(\mu) P_\ell(\mu)\,\dd\mu$
are the coefficients in a series in terms of the Legendre polynomials
$P_\ell(\mu)$, and $N_\ell$ is the truncation;
see \cite{Rae73} or \cite{KR80} for details.
Using $N_\ell=500$, the result for $2\meanA_\phi\meanB_\phi$ is given
in \Fig{pcalc_bl}, where we show its latitudinal dependence either
for the combined data set of CR~2161--2163 or for each of the three
CRs separately.
Here, $\hatB_\phi$ has been truncated to the same level using an equation
analogous to \Eq{pcalc_bl}, but with expansion coefficients computed
with $P_\ell^1(\mu)$ instead of $P_\ell(\mu)$.

It turns out that at $-15\degr$ latitude the magnetic
helicity density is mostly negative, but at $+15\degr$ latitude
it is more noisy and with positive values only for CR2163 and
mostly negative values for CR2162.
The inset shows the full latitudinal extent
and a comparison with the corresponding Cartesian result.
Here we have applied a low-pass filter with $|k_z|\leq0.1\Mm^{-1}$,
corresponding to $N_\ell=70$.
The Cartesian result agrees with the spherical harmonics reconstruction.
Both show large positive contributions throughout the southern hemisphere.
Those are caused by the systematic presence of radial fields
($\meanB_r\approx-5\G$) at $50\degr$ to $80\degr$ southern latitudes.

The values of $2|\meanA_\phi\meanB_\phi|$ are of the order of
$1000\G^2\Mm$, which is comparable to the values of $|kH_{\rm M}(k)|$
near the maximum at $k=0.1\Mm^{-1}$; see \Fig{pphelspec_two}.
This suggests that the contributions from the
azimuthally averaged mean field are captured correctly
by our two-scale analysis.
On the other hand, there is an obvious difference between these two
approaches in that in our two-scale approach the gauge analogous to
$A_\phi(\mu=\pm1)=0$ is not adopted, but instead the integral over
$A_\phi(\mu)$ is implicitly assumed to vanish.
This can cause a difference in the $k\to0$ limit.

\section{Comparison with earlier work}

Although the use of a two-scale analysis is completely new in solar physics,
some meaningful comparison with earlier work can be made.
First, the fact that the magnetic helicity is negative in the
north and positive in the south has been known for some time
\citep{See90,PCM95,BZAZ99}, but the global analysis is now able to show
that, averaged over one or several CRs, there are hardly any sign changes
in one hemisphere, i.e., the hemispheric sign rule of magnetic helicity
is well obeyed.
The average spectral current helicity at $k=0.1\Mm^{-1}$ (a scale of
approximately $60\Mm$) is about $-100\G^2$ in the northern hemisphere;
see \Figp{pphelspec_two}{b}.
Such a statement has not been possible with conventional methods,
which were restricted to just the area of one AR.
For example the approach of \cite{ZBS16} yielded values of about
$10^3\G^2$ for AR~11158 and $10^4\G^2$ for AR~11515, both in the
southern hemisphere and at $k=0.1\Mm^{-1}$.
These values are between 10 times (for AR~11158) and 100 times (for
AR~11515, which had exceptionally large helicity) larger than the
averaged spectral current helicity found here for the entire Sun.

Most of the earlier work on magnetic helicity measurements was
restricted to the total magnetic helicity over all wavenumbers and
for a finite volume around a given AR.
Such an approach involves either time integration of photospheric
magnetic helicity injection or force-free field extrapolation,
both of which are time consuming; see \cite{Valori16} for a review.
Nevertheless, as already demonstrated by \cite{ZBS14}, such values of
total magnetic helicity are similar to the magnetic helicity density
around a given patch, for example if one assumes a volume of $3\times10^6\Mm^3$
given by the area of the patch of $(186\Mm)^2$ and a height of $100\Mm$.
For the spectrum of AR~11158, the spectral current helicity density
of $10^3\G^2$ corresponds to a mean magnetic helicity of about
$3\times10^4\G^2\Mm$; see also Figure~3 of \cite{ZBS16}.
Thus, the total magnetic helicity is $10^{11}\G^2\Mm^4$.
Since $1\G^2\Mm^4=10^{32}\Mx^2$, it corresponds to $10^{43}\Mx^2$;
this agrees with earlier estimates of the gauge-invariant magnetic
helicity for this AR, using time integration of photospheric magnetic
helicity injection \citep{VAMC12,LS12} and nonlinear force-free coronal
field extrapolation \citep{Jing12,Georg13}.

The main advantage of our approach is that it can readily be applied to
global measurements covering all longitudes and latitudes over both
hemispheres at the same time.
In that way, one can efficiently average over fluctuations, especially
in cases when there is significant cancellation.
An example of this type is AR~11515, which was an extremely complex AR with
a significant amount of magnetic helicity cancellation \citep{Wang14,Lim16}.
Moreover, even though it occurred in the southern hemisphere, the
net magnetic helicity was negative \citep{Lim16}, which was explained
by a significant amount of oppositely signed magnetic helicity at
large length scales \citep{ZBS16}.
Curiously, however, the presence of an oppositely signed magnetic helicity
at large length scales in AR~11515 of 2012 July 6 is not borne out by the
present work.
This highlights the importance of applying our new approach to longer
time series covering also a range of different phases of the solar cycle.

\section{Conclusions}

The present work has shown that it is possible to generalize the notion
of a helicity spectrum to the case where the helicity is locally
modulated in a large-scale fashion, which may even include a sign change.
This approach is particularly useful for characterizing the spectrum of
solar magnetic helicity, with the aim of being able to find out whether
there is evidence for a bihelical spectrum.
Bihelical spectra have been seen in turbulent dynamo simulations where
turbulence is driven by a helical forcing function.
Surprisingly, our present results suggest that, for the Sun, the helicity
spectrum is not bihelical -- at least not at the surface.
The reason for this is not understood at present.
As we have shown in \Fig{pcalc_bl}, the contribution from the azimuthally
averaged mean magnetic field is of comparable magnitude to that from our
two-scale analysis, but it is very noisy and has only in the southern
hemisphere at $-15\degr$ significant negative contributions, which would
be in agreement with \cite{PP14} and perhaps suggestive of a
bihelical field.
However, this contribution is overwhelmed by much stronger contributions
of opposite sign at latitudes south of $-40\degr$.
It might well be that our time frame was unfortunate and that the
net magnetic helicity of the large-scale field was close to zero.
Another possibility is that the weak contribution of large-scale fields
with opposite sign is just a surface effect.
Observations of the magnetic field in the solar wind have indicated
the presence of a bihelical spectrum \citep{BSBG11} and that the signs
of the two contributions at large and small scales are the other way around
than what is expected inside the Sun.
One is therefore led to consider the possibility of the spectrum
having changed along the way since it left the Sun, which is indeed
what turbulence simulations \citep{WBM11} and mean-field models
\citep{Bon16} have now shown.
The role of the surface has yet to be studied in that respect, but there
is clearly now a need to consider theoretical models of global
convection-driven dynamos and to apply the two-scale approach to different
layers: the surface, the interior of the convection zone, and an
outer coronal layer.

\acknowledgements
We thank the referee for useful suggestions and
Yang Liu (Stanford University) for providing us with
the synoptic vector magnetograms used in the present paper.
This work utilizes {\em SOLIS} data obtained by the NSO Integrated Synoptic
Program (NISP), managed by the National Solar Observatory, which is
operated by the Association of Universities for Research in Astronomy
(AURA), Inc.\ under a cooperative agreement with the National Science
Foundation.
{\em SDO} is a mission for NASA's Living With a Star program.
This work has been supported in parts by the Swedish Research Council
grant No.\ 621-2011-5076 and the Research Council of Norway under the
FRINATEK grant No.\ 231444.

\appendix
\section{Derivation of Equations (17) and (18)}
\label{Deriv}

The purpose of this appendix is to present the derivation of
\Eqs{2EMK0kz}{kHMK0kz}.
The Fourier transform of $\BB$ yields
\EQ
\hatBB(k)=\half(\delta_{k\,k_1^-}+\delta_{-k\,k_1^-},\;
\delta_{k\,k_1^+}+\delta_{-k\,k_1^+},\;0).
\EN
We recall that in 1D, the integrals in \Eqs{2EKk}{kHKk}
still extend over positive and negative values of $k$.
Thus, the surface integral reduces to a sum of two contributions,
one with positive $k$ and one with negative $k$.
For $K=0$, \Eq{2EKk} yields
\EQA
2\tilde{E}_{\rm M}(0,k)&=&
\left.\quarter(\delta_{k'\,k_1^+}+\delta_{-k'\,k_1^+})\right|_{k'=-k} \cr
&+&\left.\quarter(\delta_{k'\,k_1^-}+\delta_{-k'\,k_1^-})\right|_{k'=k} \cr
&=&\half\left(\delta_{k\,k_1^+}+\delta_{k\,k_1^-}\right),
\ENA
which is in agreement with \Eq{2EMK0kz} and \Fig{psingle3}.
Here, the $1/4$ factor is the energy of each Fourier peak,
but there are two of them at $k=\pm k_1^-$ and at $k=\pm k_1^+$,
which explains the $1/2$ amplitudes of each of the peaks in energy.

Next, we calculate $k\tilde{H}_{\rm M}(K_0,k)$.
Since $\hatkk$ has only a $z$ component, \Eq{kHKk} yields
\EQA
k\tilde{H}_{\rm M}(K_0,k)
&=&\left.\ii\hat{k}'[\tilde{M}_{xy}(K_0,k')-\tilde{M}_{yx}(K_0,k')]\right|_{k'=-k}\quad\cr
&+&\left.\ii\hat{k}'[\tilde{M}_{xy}(K_0,k')-\tilde{M}_{yx}(K_0,k')]\right|_{k'=k}\quad
\label{kHMxyMyx}
\ENA
To compute $\tilde{M}_{xy}(K_0,k)$
and $\tilde{M}_{yx}(K_0,k)$, we need
the Fourier transforms shifted by $\pm K_0/2$.
Those are given by
\EQ
\hatBB(k+\half K_0)=\half\pmatrix{
\delta_{k\,k_1-K_0}+\delta_{-k\,k_1}\cr
\delta_{k\,k_1}+\delta_{-k\,k_1+K_0}\cr
0},
\EN
\EQ
\hatBB(k-\half K_0)=\half\pmatrix{
\delta_{k\,k_1}+\delta_{-k\,k_1-K_0}\cr
\delta_{k\,k_1+K_0}+\delta_{-k\,k_1}\cr
0}.
\EN
Thus $\tilde{M}_{xy}(K_0,k)=\quarter\delta_{-k\,k_1}$
and $\tilde{M}_{yx}(K_0,k)=\quarter\delta_{k\,k_1}$.
Therefore, \Eq{kHMxyMyx} yields
\EQA
k\tilde{H}_{\rm M}(K_0,k)
&=&\quarter\left.\ii\hat{k}'\delta_{-k'\,k_1}\right|_{k'=-k}
-\quarter\left.\ii\hat{k}'\delta_{k'\,k_1}\right|_{k'=k}\cr\cr
&=&-\half\ii\,\sgn(k_1)\,\delta_{k\,k_1}.
\ENA
which agrees with \Eq{kHMK0kz}.

%r e f

\vfill\bigskip\noindent\tiny\begin{verbatim}
$Header: /var/cvs/brandenb/tex/nishant/LShelicityspec/paper.tex,v 1.128 2017/01/24 14:40:54 brandenb Exp $
\end{verbatim}

\end{document}